\documentclass[%
 aip,
 jmp,%
 amsmath,amssymb,
 reprint,%
]{revtex4-1}

\usepackage{graphicx}% Include figure files
\usepackage{dcolumn}% Align table columns on decimal point
\usepackage{bm}% bold math

\begin{document}

\preprint{AIP/123-QED}

\title{Dynamics of the stratification process in drying colloidal dispersions studied by terahertz time-domain spectroscopy}

\author{J.L.M. van Mechelen}\email{dook.vanmechelen@ch.abb.com}
\affiliation{ABB Corporate Research, Segelhofstrasse 1K, 5405 Baden-D\"attwil, Switzerland}

%\title[Sample title]{Sample Title}% Force line breaks with \\
\thanks{The author thanks Dirk van der Marel for the use of the TeraView spectrometer, Louis van Mechelen for manufacturing the sample holder, D. Meier and Ch. Sidler (BASF Coating Services AG, Switzerland) for providing the Glasurit paints and technical support, and Xun Gu for fruitful discussions and a careful reading of the manuscript.}

\begin{abstract}
We present an optical study that reveals the bulk dynamics of the stratification process in drying colloidal dispersions. Terahertz time-domain spectroscopy has been used to measure \textit{in situ} solventborne and waterborne paint layers as a function of drying time. The dynamic behavior of the dry top layer and wet bottom layer thickness, as well as the bulk thickness, reflect the principal processes of the established drying mechanism. In addition, the results demonstrate stratification only when the drying process is in the evaporation controlled regime, whereas the coating is shown to remain a single layer for diffusion controlled drying.
\end{abstract}

\maketitle %% required

\section{Introduction}
The drying mechanism of colloidal dispersions is of widespread scientific and industrial interest, and decades of study have provided
a general understanding of the physical processes that occur in several separate regimes.~\cite{Vanderhoff,Blandin,Croll,Keddie,Keddie_book,Routh} For aqueous dispersions, we can distinguish at least three stages. In stage I, the so-called constant rate phase, water evaporates from the  coating surface, with a rate that is close to that of pure water,~\cite{Croll} and particles order. The surface water is provided from the bulk by diffusion which is as least as fast as to keep up with the evaporation rate. This process ends when the amount of water in the system drops below a critical value.~\cite{Croll,Keddie} In stage II, the falling rate phase, evaporation is controlled by diffusion of interstitial water between the particles. The lower amount of water in the system and the slowing down diffusion lead to a decreasing evaporation rate. In this stage, particles get into contact with each other and are deformed.~\cite{Keddie} In stage III, the volume fraction of water becomes small, evaporation becomes almost negligible, and the solvent diffuses across particle-particle boundaries. Eventually, particles compact together into a final dense layer.

Most processes of the drying mechanism are related to exterior signatures of the drying coating. Solvent emission and gravimetrical measurements, for instance, have provided knowledge about evaporation and diffusion.~\cite{Vanderhoff,Croll,Blandin}  The bulk nature of the drying film, on the other hand, is largely unexplored mainly due to the viscous and opaque states of matter which exclude many experimental techniques. For a long time, it has been suggested that the evaporation and diffusion characteristics are related to stratification during film formation.\cite{Sheetz,Croll} Recent calculations show that a dry top layer should form on the wet bulk when the rate of evaporation is larger than that of diffusion.~\cite{RouthZimmerman} This prediction has obtained experimental support from \textit{in situ} nuclear magnetic resonance (NMR) data which suggest solvent depletion in the upper part of the coating layer.~\cite{Gorce,Ciampi} More recently, \textit{ex situ} scanning electron microscopy (SEM) has shown particle agglomeration in the upper region of a latex coating at a given state of drying~\cite{Ma,Luo}. Additional insight into the mechanism could be gained by probing the thickness evolution of the presumed layers. Previous work, however, has shown that it is even not trivial to measure \textit{in situ} the total bulk thickness. Optical techniques using visible and infrared light and NMR have deduced the total thickness evolution during drying\cite{Gorce,Ciampi,Goossens,Martinez,Sargent,Yucel,Unsal,Ludwig} although these data often are noisy, have a poor resolution, and are not directly measured but model determined. A precise determination of the layered behavior during film formation requires a technology which probes \textit {in situ} the bulk properties in a noncontact manner and is sensitive to the presence of different layers.

Terahertz spectroscopy is an upcoming technology, and, although mostly used in physics, its inherent property of depth sensitivity for many materials extends its application into other fields such as medicine and biology~\cite{Tonouchi}. Also fundamental issues in chemistry are increasingly studied with this low energetic radiation which often tunes to the vibration and rotation degrees of freedom. It was shown that terahertz radiation partially penetrates wet solventborne paint layers during drying.~\cite{Yasui2005} However, significant approximations of the analysis applied on data with low signal-to-noise ratios provided relative inaccurate bulk thickness estimates.~\cite{Yasuda2007,Cook2008,Iwata}

Precise material properties of solids can be obtained from terahertz data using dedicated analysis methods.~\cite{VanMechelen2014} Chemistry often deals with liquidlike samples, for which traditionally spectroscopic peak analyses are performed. In order to determine the macroscopic properties of chemical species, no standard model approach exists, and sometimes methods are applied as for solids~\cite{Vinh}.

In this work, we have studied the bulk drying behavior of different colloidal dispersions with terahertz time-domain spectroscopy between the as-deposited and fully dry state. By using an analysis approach based on a stratified dispersive model, we obtain simultaneously the optical and geometrical properties of the coatings with high accuracy. We show that stratification occurs depending on the relative rate of evaporation as compared to diffusion, as theoretically predicted. We discuss the geometrical properties of the drying coating in the context of the established three stage drying mechanism.

\section{Experiment and methods}
The process of film formation is of importance both from a fundamental as well as from an application point of view. For this reason, our studied samples reflect the situation as encountered in the automotive painting industry. Here, a typical multilayer coating may consist of primer, base coat, and clear coat, which have individual dry layer thicknesses between $10-50$\,$\mu$m, and the total stack is below 100\,$\mu$m. The paints are either organic or aqueous colloidal dispersions, and are typically referred to as solventborne and waterborne, respectively. Apart from the clear coat, all paints contain in addition to the binder and solvent also pigment (see Table~\ref{table1}). The constituents of the waterborne base coats are given in Table~\ref{table2} (as obtained from BASF).

\begin{table}[tb]
\begin{center}\caption{\label{table1}\textbf{Basic properties of the used automotive paints (Glasurit line, BASF).}}
\begin{tabular}{l l l l }
\hline
paint\quad\qquad\ &binder\qquad\qquad\ &solvent\qquad\quad  &product nr. \\ \hline
primer & epoxy resin & organic &  801-72 \\
base coat & polyurethane & organic &  22-VOC-3.5 \\
base coat & polyurethane & water &  RH-90  \\
clear coat & saturated polyester & organic  & 923-135 \\
 & thermosetting &  & \\
 & polyacrylate resin &   &  \\ \hline
\end{tabular}
\end{center}
\end{table}
\begin{table}[tb]
\begin{center}\caption{\label{table2}\textbf{Constituents of waterborne base coats (in wt \%).}}
\begin{tabular}{l c c c c c}
\hline
paint\qquad\ &solid &solvent\textsuperscript{\emph{a}}\qquad  & water & pigment\textsuperscript{\emph{b}} & particles\textsuperscript{\emph{b,c}}\\ \hline
plain blue& 26.4 & 12.4 &  61.2 & 10.1 & -\\
mica black & 15.8 & 11.0 &  73.2 & 0.8 & 0.01 Al, 0.08 mica\\ \hline
\end{tabular}
\end{center}
\vspace{-2.5mm}\textsuperscript{\emph{a}}Solvent indicates the organic solvent; \textsuperscript{\emph{b}}Pigment and particles are part of the solid content; \textsuperscript{\emph{c}}the nominal size of which 50\,\% is smaller $d_{50}=50\,\mu$m.
\end{table}

We have air-sprayed sequences of paint layers on $25\times25$ mm$^2$ optically flat steel substrates. After deposition, the samples are transferred to a (pre-)heated sample holder at $40\,^\circ$C, where they are clamped at the edges, such that they face the terahertz beam from above. The beam is laterally centered on the sample which avoids sensing areas of inhomogeneous drying. The terahertz beam has a waist $w$ with $2w=0.25-3$ mm, depending on frequency, and an average beam power of about 10\,$\mu$W. Measurements were performed by terahertz time-domain spectroscopy in the range $0.03-3$ THz (TPI spectra 1000, TeraView Ltd.) in reflection mode at an angle of incidence of about 15$^\circ$ in ambient air. Data have been recorded 20-25\,s after deposition at 30 Hz with 200 averages, resulting in an acquisition time of 7.1\,s per measurement, including internal data processing.

We have previously reported about a novel terahertz material analysis approach which allows for high-precision material parameter determination.~\cite{VanMechelen2014} The concept of the model is first to realistically describe the light-matter interaction with the sample based on general material characteristics, then to calculate the light propagation through the system and subsequently to obtain the precise material properties by performing a fit to the experimental data in the time-domain and frequency-domain.

It is advantageous to describe the studied material in its environment (e.g., air and substrate) as a stratified system. For each layer $k$, the dielectric dispersion is modeled by using oscillators that represent the physical processes that occur upon the light-matter interaction. For common materials, such as (dry) paint, these processes are generally limited to lattice vibrations and (free and/or collective) electron oscillations, which can be often well described by Lorentzian line shapes in the optical functions. This summation of Lorentz oscillators is known as the Drude-Lorentz parametrization and can be written for dielectric function $\epsilon(\omega)$ as
\begin{equation}\label{DrudeLorentz}
  \epsilon(\omega)=\epsilon_{\infty}+\sum_{\ell=1}^m \frac{\omega_{p,\ell}^2}{\omega_{0,\ell}^2-\omega^2-i\gamma_\ell\omega},
\end{equation}
where $\epsilon_{\infty}$ is the high frequency limit of $\epsilon(\omega)$, $\omega_{p,\ell}$ the plasma frequency, $\omega_{0,\ell}$ the characteristic frequency, and $\gamma_\ell$ the relaxation rate of excitation $\ell$.

The light propagation through the stratified system can be calculated with the Fresnel equations and by using the dielectric dispersion characteristics of each layer (Eq.~\ref{DrudeLorentz}). The reflected electric field $E_r$ is related to the incident electric field, $E_{r,0}$ by the transfer function $\mathcal{T}$ through
\begin{equation}\label{fresnel1}
E_{r}(\omega)=\mathcal{T}(\omega)E_{r,0}(\omega)
\end{equation}
where $\mathcal{T}$ is composed of the summation of all reflection $r_{ij}$ and transmission coefficients $t_{ij}$  corresponding to the partial reflections and transmissions on all interfaces of the system,
\begin{equation}\label{fresnel2}
\begin{split}
 \mathcal{T}(\omega) = & \ r_{12}+t_{12}r_{23}t_{21}e^{-i2\beta_2}+t_{12}r_{23}r_{21}r_{23}t_{21}e^{-i4\beta_2}\\
 & +t_{12}t_{23}r_{34}t_{32}t_{21}e^{-i2(\beta_2+\beta_3)}+\ldots
\end{split}
\end{equation}
with
\begin{equation}\label{fresnelcoeff}
 t_{ij}=\frac{2n_i}{n_i+n_j},\quad\qquad r_{ij}=\frac{n_i-n_j}{n_i+n_j}
\end{equation}
where $\beta_k=\omega n_k d_k/c$ is the phase shift accumulated in layer $k$, $\omega$ is the angular frequency of the terahertz radiation, $n_k=\sqrt{\epsilon_k}$ is the complex index of refraction representing the optical properties of layer $k$, $d_k$ is the thickness of layer $k$, and $c$ is the speed of light in vacuum.

A typical reflection measurement probes the reflected electric field of the sample $E_r^{\text{exp}}(t)$ and of a reference $E_{r,0}^{\text{exp}}(t)$ which contains the characteristics of the terahertz pulse shape and the experimental setup.  For the reference,  a material is chosen with known optical properties, typically a metal due to the high and almost frequency independent reflectivity, commonly $R(\omega)>99.5\,\%$ at 0.3 THz. In case the sample configuration is more complicated than a single layer, a  convenient method to obtain the material properties $d_k$ and $\epsilon_k$ is by applying a fitting approach to the time-domain functions $E_r^{\text{exp}}(t)$ and $E_{r,0}^{\text{exp}}(t)$. However, inherent to the least-squares algorithm, optimization is mainly done for the largest values of $E_r^{\text{exp}}(t)$ which does not correspond to fitting all probed frequencies equally. A better result can therefore be obtained by performing simultaneously a fit in the frequency-domain to a Fourier transformed function of $E_r^{\text{exp}}(t)$. For the latter, a convenient set is the real and imaginary part of the reflectivity $r(\omega)=E_r(\omega)/E_{r,0}(\omega)$, since values are typically well spread  between the bounds, $-1$ and $+1$, unlike $R(\omega)=|r(\omega)|^2$ which is often close to 1.

A main advantage of this analysis approach is that the material properties can be realistically described which results in highly accurate fit parameters without prior knowledge of either the specific optical functions or the thicknesses. The moisture in the measurement environment is taken into account either directly by $E_{r,0}^{\text{exp}}(t)$ in case this is measured at the same humidity level, or by modeling the air layer with the use of oscillators that characterize water absorption in the terahertz range.

\section{Results and discussion}
In the first part of this section we show how to obtain the optical and geometrical properties of polymeric coatings in the initial, as-deposited, and the final, fully dry states. In the second part, we use these optical properties to analyze the drying coatings at all states in between the as-deposited and fully dry state, which provides the dynamics of their geometrical properties. In the third part, the experimental evidence on the occurrence of stratification and the balance of evaporation and diffusion is put in the context of theoretical predictions.

\begin{figure}[htbp]
\centerline{\includegraphics[width=\columnwidth]{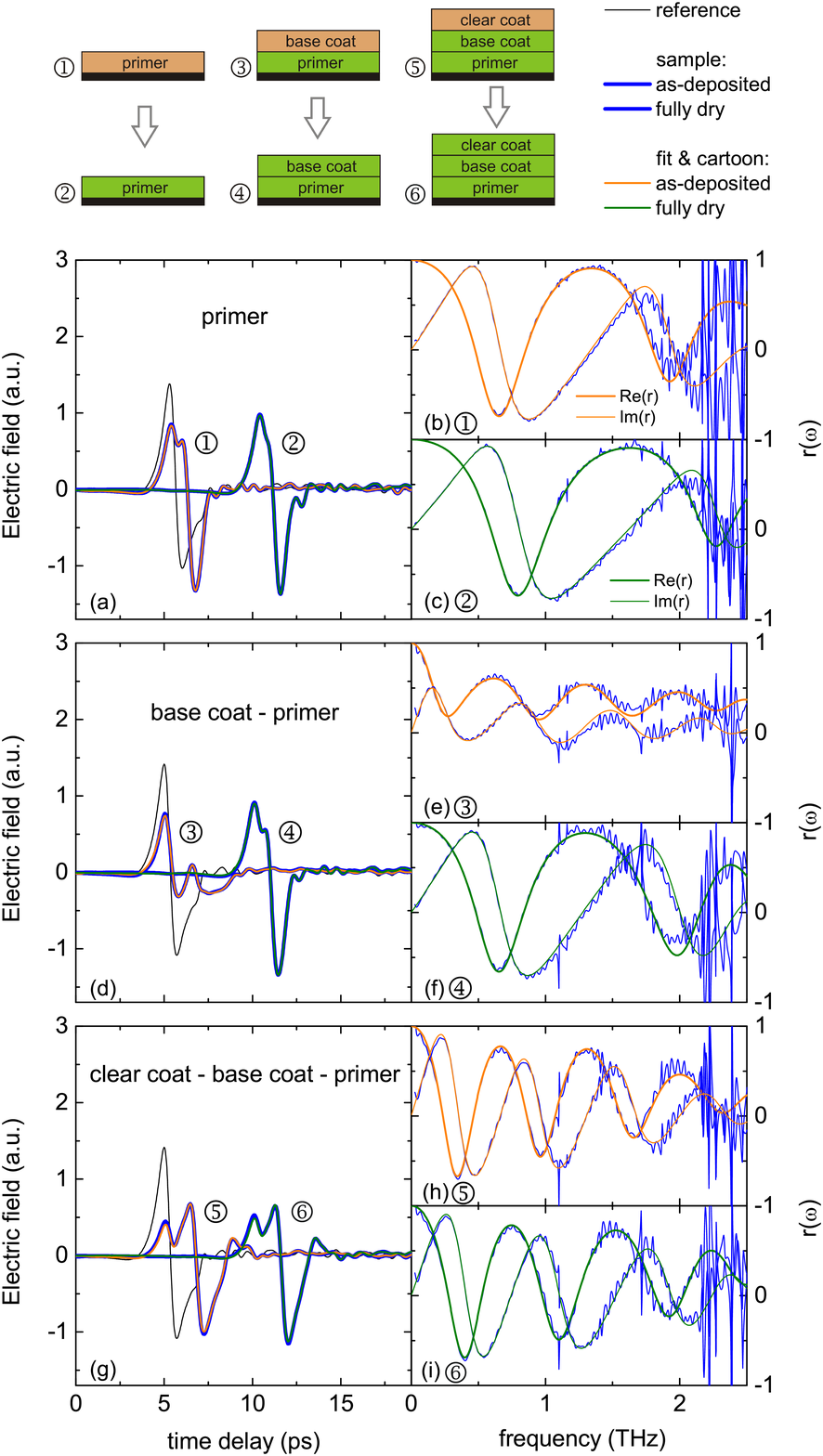}}
\caption{$E_{r}^{\text{exp}}(t)$, Re\,$r^{\text{exp}}(\omega)$, Im\,$r^{\text{exp}}(\omega)$ and fits $E_r(t)$, Re\,$r(\omega)$, Im\,$r(\omega)$ of an incrementally deposited paint stack shown at each as-deposited and fully dry state, as illustrated in the cartoon image, and $E_{r,0}^{\text{exp}}(t)$ of a steel substrate. The as-deposited state always refers to the upper layer of the stack, whereas the dry state refers to the entire stack. (a-c) \textit{primer--steel} (d-f) \textit{base coat--primer--steel} (g-i) \textit{clear coat--base coat--primer--steel}.  $E_r^{\text{exp}}(t)$ have been temporally shifted in order to have sample and reference geometrically aligned. For clarity, $E_{r}^{\text{exp}}(t)$  of dry stacks have been shifted by an additional +\,5 ps. Experimental conditions: 40\,$^\circ$C and $50\pm4$\,\% relative humidity of the $25\pm2\,^\circ$C ambient air.}
\label{fig1}
\end{figure}

\subsection{As-deposited and fully dry state}
Fig.~\ref{fig1} shows $E_r^{\text{exp}}(t)$ of a paint layer stack at three incremental stages of deposition. After deposition of the first layer (\textit{primer}), the sample is measured, then dried and measured again. This provides $E_r^{\text{exp}}(t)$ in the as-deposited and fully dry state (see Fig.~\ref{fig1}a). Subsequently, a next layer has been deposited on the dried stack and the measurement and drying procedure is repeated (see Fig.~\ref{fig1}d,g). After three depositions, the stack consists of \textit{clear coat -- base coat -- primer -- steel}, where the base coat is waterborne mica black (see Table~\ref{table2}). The reference $E_{r,0}^{\text{exp}}(t)$ has been measured on a pristine steel substrate (Fig.~\ref{fig1}a,d,g). The relative position of the paint surface as compared to the reference surface, $x$, varies per measurement due to shrinkage of the paint and repositioning of the stack after each deposition, but is at most 1 mm and thus smaller than the Rayleigh range of the terahertz beam. In order to mutually compare the curves in Fig.~\ref{fig1}, the displayed $E_{r}^{\text{exp}}(t)$ have been temporally displaced in order to cancel the variation of $x$, obtained from the analysis (see below). The right panels show for each stack the corresponding Re\,$r^{\text{exp}}(\omega)$ and Im\,$r^{\text{exp}}(\omega)$ in the frequency-domain, using the displaced $E_{r}^{\text{exp}}(t)$. The reflected response changes significantly between the as-deposited and dry state for primer (Fig.~\ref{fig1}a-c) and base coat (Fig.~\ref{fig1}d-f), and less for clear coat (Fig.~\ref{fig1}g-i). The noise on $r^{\text{exp}}(\omega)$ is mainly caused by the presence of polar solvents in the paint, and is enhanced for $\omega<2$\,THz for waterborne paint which contains more than 70\,\% water in the as-deposited state.

\begin{table}[b]
\begin{center}\caption{\label{table3}\textbf{Thickness of the upper layer of the incremental paint stack (see Fig.~\ref{fig1}) from fits (in $\mu$m).}}
\begin{tabular}{l l l l}
\hline
state\qquad\qquad\ &primer\qquad\quad\ &base coat\qquad\  &  clear coat \quad\ \\ \hline
as-deposited & $52.6\pm0.4$ & $66.7\pm0.6$ & $68.4\pm3.2$\\
fully dry & $40.3\pm0.2$ & $11.6\pm0.2$ &  $50.8\pm1.9$\\ \hline
\end{tabular}
\end{center}
\end{table}

We have previously shown that dry paint multilayers can be characterized using an analysis approach that is based on a stratified dispersive model~\cite{VanMechelen2014}, as described in section \textit{Experiment and methods}. A dry paint layer consists of one or several polymer macromolecules that fully span the sample size which is much larger than $\lambda=c/\omega$, and can therefore be treated as homogeneous for terahertz radiation. In order to apply this approach also to wet paint, it is important to consider at each drying stage  the size of the optical domains relative to $\lambda$, since comparable sizes leads to diffraction which is not included in the model. As-deposited paint is a colloidal dispersion composed of resin (polymer), pigments, solvents, and additives. The majority of particles is from the pigment which have diameters below 0.5\,$\mu$m $\ll\lambda$. The solvent, on the other hand, forms a large domain which spans the sample size~\cite{Ma} and is therefore much larger than $\lambda$. For these reasons, we treat as-deposited paint as homogeneous with average optical properties of solid and liquid material. During the drying process, particles agglomerate and coalesce and may form domains with sizes that are comparable to $\lambda$ before reaching the fully dry state. We estimate, however, that this region of phase space where scattering may occur is rather small, and we treat throughout this work also the drying coating layer as homogeneous for terahertz radiation.

For dry paint layers, the dielectric dispersion (see Eq.~\ref{DrudeLorentz}) can be modeled with one or two Lorentz oscillators which represent the optical behavior at higher (infrared) frequencies and, if present, phonon resonances due to for example pigments.~\cite{VanMechelen2014}  Although as-deposited paint is liquid-like, we show that it can be modeled with a similar dielectric dispersion as for dry paint. Note that the dielectric relaxation of water molecules in waterborne paint requires an additional oscillator at gigahertz frequencies. We therefore analyze the as-deposited state similarly as the dry state and model the entire system as an \textit{air -- paint stack -- substrate} multilayer, where the \textit{paint stack} itself is also a multilayer corresponding to the individual paint layers.

Fig.~\ref{fig1} shows the result of applying the stratified dispersive model to the incremental paint layer stack in the as-deposited and fully dry state. The quality of the time-domain and frequency-domain fits demonstrates that besides dry paint also as-deposited paint can be well described by the model. The thickness of the upper paint layer is given in Table~\ref{table3} and illustrates the shrinkage from the as-deposited to the dry state, which is considerable for waterborne paints due to evaporation, as discussed later. There is currently no commercial measurement technique available to compare the thickness values in the as-deposited state. The total layer thickness in the dry state has been measured by well-established magnetic induction, which gives $102\pm3\,\mu$m versus $102.7\pm0.5\,\mu$m determined in this work for the stack \textit{clear coat -- base coat -- primer -- steel}, corresponding to Fig.~\ref{fig1}g-i.

\begin{figure*}[htbp]
\centerline{\includegraphics[width=0.92\textwidth]{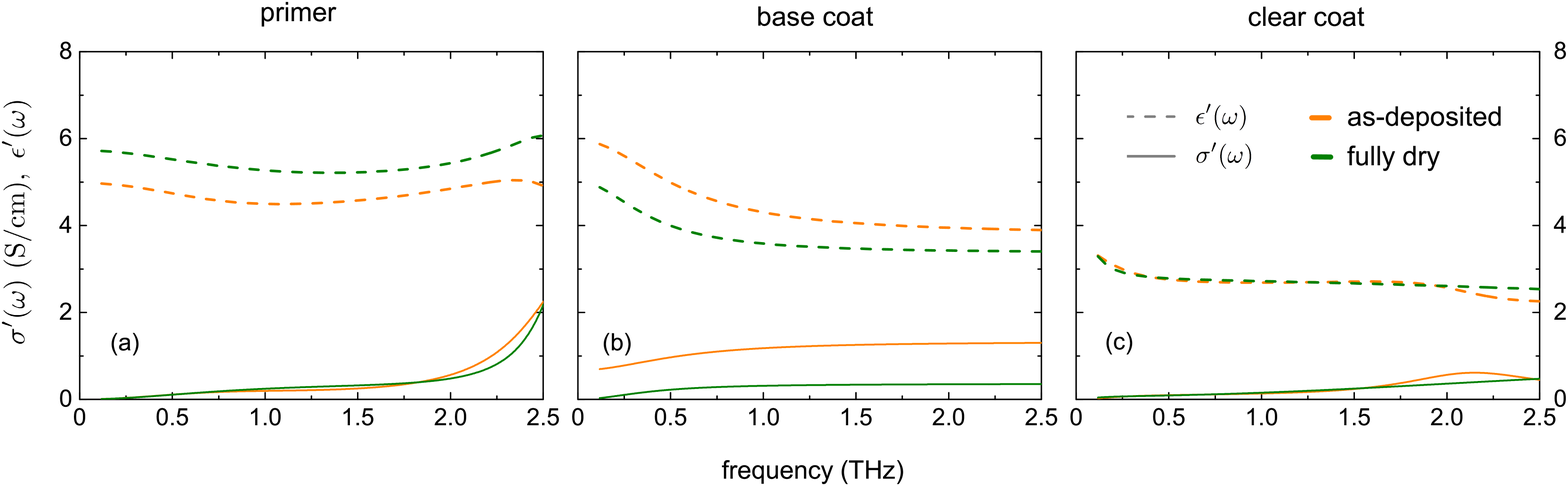}}
\caption{Real part of the dielectric function $\epsilon'(\omega)$ and real part of the optical conductivity $\sigma'(\omega)$ of (a) solventborne primer, (b) waterborne mica black base coat, and (c) solventborne clear coat, in the as-deposited and fully dry state.}
\label{fig2}
\end{figure*}

Fig.~\ref{fig2} shows the optical properties of the upper layer at each incremental stage, as manifested by the dielectric function $\epsilon(\omega)=\epsilon'(\omega)+i\sigma'(\omega)/\epsilon_0\omega$ obtained from the analysis, where $\sigma'$ is the real part of the optical conductivity. Note the very different behavior of the various paints upon drying. For solventborne primer (Fig.~\ref{fig2}a), $\epsilon'$ increases by about 20\,\%, whereas $\sigma'$ almost does not change. The upturn of $\sigma'$ toward higher frequencies is most probably caused by phonon absorption from the pigment. The terahertz properties of waterborne mica black base coat (Fig.~\ref{fig2}b) are drastically influenced by the drying, presumably due to the diminishing spectral weight from the dielectric relaxation of water. Optically transparent clear coat (Fig.~\ref{fig2}c) changes only marginally during drying. The reason of the slightly enlarged conductivity around 2 THz, as well as the low frequency behavior is unknown. Comparison between Fig.~\ref{fig2} and Table~\ref{table3} shows that the relative change of the optical properties of these automotive paints may be as large as the geometrical shrinkage. This relates back to earlier work on paint thickness determination using terahertz radiation,~\cite{Yasuda2007,Cook2008,Iwata} and demonstrates that the optical properties may not be considered invariant as a function of both drying time and frequency.

\subsection{Stratification process}
\begin{figure*}[htbp]
\centerline{\includegraphics[width=\textwidth]{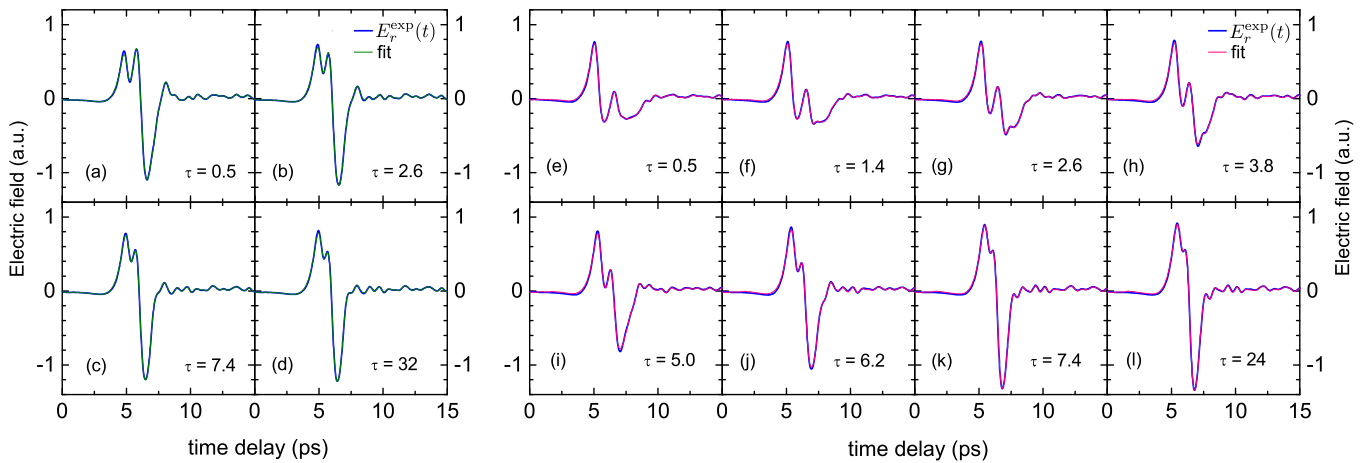}}
\caption{$E_r^{\text{exp}}(t)$ and fits $E_r(t)$ of (a-d) \textit{primer -- steel} and (e-l) \textit{black mica base coat -- primer -- steel}, where only the upper layer is wet, as a function of drying time $\tau$ (in min) at 40\,$^\circ$C and $53\pm2$\,\% relative humidity of the $25\pm2\,^\circ$C ambient air. Curves in panels (e,l) are identical to those in Fig.~\ref{fig1}d.}
\label{fig3}
\end{figure*}

We now concentrate on the drying process, and study the paints in between the as-deposited and fully dry state with the goal to determine the geometrical properties during drying. In the following we use the designation `wet' for any state from as-deposited until fully dry. Fig.~\ref{fig3} shows $E_r^{\text{exp}}(t)$ of wet solventborne primer on steel (a-d) and wet waterborne base coat on dry primer on steel (e-l) for selected values of the time $\tau$ elapsed after deposition. In contrast to Fig.~\ref{fig1}, the displayed curves are not corrected for the varying air layer $x$. $E_r^{\text{exp}}(t)$ of the primer, which is thicker than the sample shown in Fig.~\ref{fig1}a, changes both due to shrinkage and the increase of $\epsilon'$ in comparable amounts. For the base coat, it is the very large shrinkage (cf.\ Table~\ref{table3}) which dominates the overall behavior of $E_r^{\text{exp}}(t)$.

As mentioned in the Introduction, theoretical and experimental work suggest that the drying mechanism of colloidal dispersions may involve stratification. The depth sensitivity of terahertz spectroscopy together with the analysis approach, as used before for as-deposited and fully dry paint stacks, are particularly suited to investigate the stratification process with high precision. We concluded earlier that the drying coating layer can be considered as homogeneous for terahertz radiation in most of the phase space. Based on these considerations, we have analyzed $E_r^{\text{exp}}(t)$, Im\,$r^{\text{exp}}(\omega)$ and Re\,$r^{\text{exp}}(\omega)$ of the two paint stacks using the stratified dispersive model where the wet paint is modeled as a double layer. The top layer has the optical properties of dry paint and the bottom layer of as-deposited paint, both obtained from previous analyses (see section \textit{Results and discussion, as-deposited and fully dry state}). The three fitting parameters are the thickness of the air above the paint stack $x$, the thickness of the top layer $d_{\text{top}}$, and the thickness of the bottom layer $d_{\text{bottom}}$.

\begin{figure}
\centerline{\includegraphics[width=0.8\columnwidth]{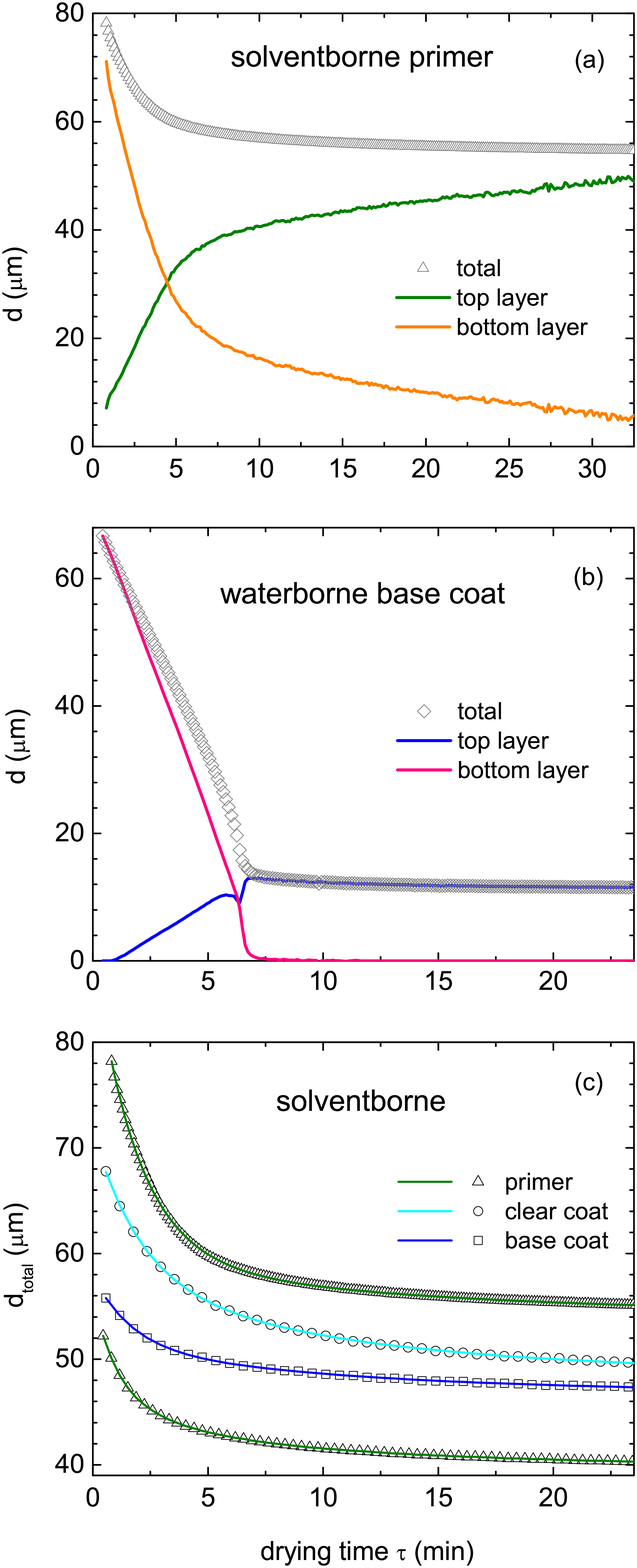}}
\caption{Layer thicknesses $d$ of the top layer $d_{\text{top}}$, the bottom layer $d_{\text{bottom}}$, and their total $d_{\text{total}}$ as a function of drying time $\tau$ of (a) solventborne primer and (b) waterborne mica base coat. Panel (c) shows $d_{\text{bottom}}$ for a larger set of solventborne paints as well as the fit results using a double exponential function (lines).}
\label{fig4}
\end{figure}

Fig.~\ref{fig3} shows the result of the time-domain fits for the primer and base coat. The fits match well the experimental data at all $\tau$, also for waterborne base coat where the changes of $E_r^{\text{exp}}(t)$ are significant. The used approach of a bilayer system with hard boundaries does not take into consideration the small region in between the two layers where previous studies show a concentration gradient~\cite{Gorce,Ciampi,Ma}. A blurred interface between two layers shows up in $E_r^{\text{exp}}(t)$ as a broadened structure as compared to a discontinuous interface. The perfect match between the data and the model in Fig.~\ref{fig3} suggests that the optical thickness of the gradient region is small as compared to the optical thickness of the entire layer, and that a bilayer model applies well in the current case. The fitting parameters $d_{\text{top}}$ and $d_{\text{bottom}}$ and $d_{\text{total}}=d_{\text{top}}+d_{\text{bottom}}$ are presented in Fig.~\ref{fig4}(a,b). We first discuss the results of solventborne primer followed by waterborne base coat, and relate them to the drying mechanism of each paint.

Solventborne primer manifests a double layer structure, already $\sim0.5$ min after deposition (see Fig.~\ref{fig4}a). A thin dry layer grows linearly on top of a bulk wet layer which shrinks in an exponential way, such that $d_{\text{total}}$ also decreases exponentially. Linear extrapolation of $d_{\text{top}}(\tau)$ for $\tau\rightarrow0$ suggests that stratification already started right after deposition. Although the linearity of $d_{\text{top}}(\tau)$ tends correlation with the linear evaporation rate of water in latex paints, the initial exponential character of $d_{\text{total}}$ shows that the process at this stage is not fully evaporation controlled. In these solventborne paints, preformed low molecular weight polymers undergo cross-linking which is of main importance for the film formation, and processes similar to the ones occurring in stage I of aqueous dispersions (see \textit{Introduction}) can be either reduced or absent.~\cite{Amalvy} For $\tau\gtrsim7$ min, the fast rate of change of all thicknesses decreases to a much slower behavior, and for $\tau\gtrsim$ 12-13 min,  $d_{\text{top}}$ and $d_{\text{bottom}}$ have a mainly linear behavior which reveals that the top layer grows slightly slower than the shrinkage of the bottom layer. For $\tau\gtrsim20$ min, $d_{\text{total}}$ has almost reached its final value, contrary to $d_{\text{top}}$ and $d_{\text{bottom}}$. Exponential extrapolation of $d_{\text{bottom}}\rightarrow0$ indicates that the drying process is only complete at $\tau\sim50$ min. Consequently, the behavior of $d_{\text{total}}(\tau)$ does not reflect the interior mechanism in the second regime, for $20\lesssim\tau<50$ min.

The decreasing rate of change of $d_{\text{total}}$ in the context of the proximity of solventborne paint to aqueous dispersions and thus to the three stage drying model suggests that film formation in solventborne primer is also characterized by evaporation and diffusion. Since both processes are expected to have distinct time scales~\cite{Routh}, we have fitted $d_{\text{total}}(\tau)$ to a double exponential function and verified that this was the smallest sum of exponentials needed for a satisfactorily fit. Fig.~\ref{fig4}c shows $d_{\text{total}}$ and the fit results for a larger set of solventborne paints. The two characteristic drying times $\tau_1$ and $\tau_2$, as well as their amplitude ratio $\mathcal{A}=A_1/A_2$ are given in Table~\ref{table4}. $\tau_1$ turns out to be characteristic for the kind of paint and increases with its volatility, whereas $\tau_2$ is large and constant for all paints. This seems to relate to the theoretical picture for aqueous dispersions where stage II, characterized by the smaller time constant $\tau_1$, is mediated by evaporation and stage III, characterized by $\tau_2$, driven by solvent diffusion which is similar for all drying sessions due to the constant temperature. The amplitude ratio $\mathcal{A}$ demonstrates the degree of evaporation as compared to diffusion, which is most pronounced for primer.

For waterborne base coat, Fig.~\ref{fig4}b shows a very different behavior than for solventborne paint. At the beginning of the measurement, $\tau\sim0.5$\,min, the result of the analysis procedure indicates the presence of a single wet layer which quickly shrinks. For $\tau\sim1$\,min, a dry layer appears on top of the wet layer, which grows with $\tau$. The minimum thickness which can be resolved by terahertz spectroscopy depends on the optical thickness and the measurement geometry and can be well below $1\,\mu$m.~\cite{Withawat} The question arises if for $\tau<1$\,min the technology cannot resolve the toplayer or if the system is a true monolayer. Since the first resolved thickness $d_{\text{top}}(\text{1 min})\approx0.11\ \mu$m is much smaller than the pigment particle size and of the same order as the polymer chains, we conjecture that for $\tau<1$\,min the system is composed of a true single layer.

It is remarkable that although $d_{\text{total}}(\tau)$ varies only approximately linearly, $d_{\text{bottom}}(\tau)$ and $d_{\text{top}}(\tau)$ are true linear functions. This fits the established opinion that the initial phase of film formation, stage I, is governed by a constant evaporation rate of water. It is interesting to observe that also the dry layer thickness increases in a linear way, which confirms the conjecture of Luo \textit{et al.}~\cite{Luo} based on \textit{ex situ} cryo-SEM studies. The ratio of the slopes of $d_{\text{bottom}}(\tau)$ and $d_{\text{top}}(\tau)$ is a measure for the conversion of wet into dry paint and should thus be comparable to the total shrinkage if evaporation is the only factor in drying. The ratio of the slopes gives 4.4 as compared to a shrinkage of 5.8 (see Table~\ref{table3}), which indicates that evaporation is the most important mechanism of film formation in latex paints. The remainder of the shrinkage, that is, 24\,\%, is supposed to be due to the initially spherical particles which deform to fill the available space. This matches well the increase of the filling fraction of close-packed hard spheres to a homogeneous layer, i.e. from 0.74 to 1, respectively, pointing to an increase of 26\,\%. Another direct correlation to previous measurements is the difference of the absolute values of the slopes of $d_{\text{bottom}}(\tau)$ and $d_{\text{top}}(\tau)$, which translates into a loss of material of $1.24\cdot10^{-5}$ kg/m$^2$s. In comparison, the evaporation rate of water at $40\,^\circ$C surrounded by still air at $25\,^\circ$C with $50\,\%$ relative humidity, is calculated to be $1.5\cdot10^{-5}$ kg/m$^2$s, which points to an evaporation rate of $83\,\%$ as compared to a pure water surface. This relates well to Croll~\cite{Croll} who concluded about 85\,\%.

After this linear phase, the slopes of $d_{\text{total}}(\tau)$ and $d_{\text{bottom}}(\tau)$ increase until both show an abrupt kink. Also Martinez \textit{et al.}~\cite{Martinez} report data where for some latex paints the thickness changes strongly, although the effect has been left uncommented. $d_{\text{top}}(\tau)$ shows a dip at the moment where the slope of $d_{\text{bottom}}(\tau)$ increases. Keddie \textit{et al.}~\cite{Keddie} showed a sudden behavior in the (surface) optical properties around the same stage of drying. The explanation for this effect was the appearance of surface roughness involving subwavelength sized voids, although the surface imaging technique environmental SEM could not capture this effect. However, since terahertz wavelengths are not sensitive for subvisible wavelength voids, this argument cannot explain the abrupt behavior of the thicknesses. It has also been suggested that the constant rate phase ends at a critical volume fraction.~\cite{Croll,Keddie} Since the state of matter at these points may drastically change, we conjecture that for $\tau$ between 5 and 7 min the double layer model where the layers have fixed optical properties is probably too simple. A fitting procedure where also the optical properties of both layers were allowed to vary, led however to undetermined geometrical and optical parameters, probably due to lack of characteristic features in the data.

With increasing drying time, $d_{\text{bottom}}$ goes rapidly to zero, and the waterborne paint can be again considered as a single layer system. $d_{\text{total}}$ gradually decreases, similar as observed for solventborne paints.  In this single-layer regime, $d_{\text{total}}(\tau)$ can be well fitted to a double exponential function which also here indicates the presence of two different time scales $\tau_1$ and $\tau_2$ due to evaporation and diffusion, respectively (see Table~\ref{table4}). It is interesting to notice that $\mathcal{A}<1$ and $\tau_1<\tau_2$ which points to a process of important but slow diffusion next to fast but little evaporation. Plain blue waterborne base coat (see Table~\ref{table2}) shows qualitatively the same signatures as the mica black base coat, including the initial region of $d_{\text{top}}(\tau)=0$, the linear behavior, the kink structure and the single layer behavior for larger $\tau$.

\begin{table}[b]
\begin{center}\caption{\label{table4}\textbf{Characteristic times $\tau_1$ and $\tau_2$ (in min) and amplitude ratio $\mathcal{A}$ of  the drying behavior of solventborne and waterborne paints obtained from $\mathbf{d_{\text{total}}(\tau)}$.}}
\begin{tabular}{l l l l l}
\hline
paint type\hspace{1.4cm}\qquad\ &$\tau_1$\qquad\qquad\ &$\tau_2$\qquad\qquad\  &$A_1/A_2$\\ \hline
solventborne primer& $1.5\pm0.3$ & $11.0\pm1.5$ & $2.8\pm1.0$\\
solventborne base coat & $1.7\pm0.3$ & $10.9\pm0.7$ & $1.1\pm0.2$ \\
solventborne clear coat& $2.1\pm0.2$ & $10.2\pm0.7$ & $1.6\pm0.4$  \\
waterborne base coat\textsuperscript{\emph{a}}& $3.0\pm0.3$ & $10.3\pm0.8$\ & $0.5\pm0.1$ \\\hline
\end{tabular}
\\ \hspace{-15mm} \textsuperscript{\emph{a}}determined from $d_{\text{total}}(\tau)$ in the exponential regime
\end{center}
\end{table}

\subsection{The balance of evaporation and diffusion}
Our terahertz data indicate that both solventborne and waterborne paints manifest stratification during film formation. Recent calculations~\cite{RouthZimmerman}  predict, however, that bilayer formation does not always occur in dispersions and depends on the ratio of the time scales of diffusion $\tau_{\text{diff}}$ and evaporation $\tau_{\text{evap}}$, the so-called P\'eclet number, $Pe=\tau_\text{diff}/\tau_{\text{evap}}$. For $Pe\gg1$, the process is mainly evaporation driven and the lack of diffusion creates a solvent depleted top layer, whereas $Pe\ll1$ indicates a diffusion controlled process where the system stays homogeneous throughout the entire process.

For solventborne paints used in this study, $\tau_{\text{evap}}$ and $\tau_{\text{diff}}$ are readily given by $\tau_1$ and $\tau_2$, respectively, which yields $Pe\approx5-7$ for the entire drying process. Film formation is thus evaporation driven and should theoretically result in the formation of a skin layer, as also the  analysis of our terahertz data has shown.

For waterborne paints, the situation is more complicated. In stage I, $\tau_{\text{evap}}$ can be calculated from the ratio of $d_{\text{total}}(0)\approx66.7\ \mu$m and the rate of evaporation.~\cite{Routh} The latter is simply obtained from the absolute value of the slope of $d_{\text{total}}(\tau)$ (see Fig.~\ref{fig4}b), yielding $\tau_{\text{evap}}\approx9$ min. The time scale for diffusion depends on the degree of particle packing which drastically increases with $\tau$. Therefore, in stage I, $\tau_{\text{diff}}<\tau_2$ which gives an upper limit of $Pe\approx1.1$ (cf.\,Table~\ref{table4}). We conjecture, however, that right after deposition the dispersion is very dilute as compared to stages II-III which leads to $Pe\ll1$ implying a diffusion driven drying process. The theoretical prediction is that no stratification should occur, as also experimentally observed for $\tau<1$ min. Evaluation~\cite{RouthZimmerman} of $\tau_{\text{diff}}$ for pure water at $40\,^\circ$C and with particles smaller than $0.3\,\mu$m, in combination with the previously calculated value of $\tau_{\text{evap}}$ support this scenario. Eventually, in stage II and III, $\tau_{\text{evap}}\approx\tau_1$ and $\tau_{\text{diff}}\approx\tau_2$ which yields $\tau_{\text{diff}}>\tau_{\text{evap}}$ and thus indicates that the process is evaporation driven and a skin layer should be formed, as observed.

The occurrence of stratification deduced from our terahertz data is thus consistent with the theoretical predictions and  depends on the balance of the evaporation rate and diffusion rate. Although in most of the phase space, the two paint classes are evaporation controlled, the slope of $d_{\text{top}}(\tau)$ relative to $d_{\text{total}}(0)$ shows that solventborne paint is more prone to stratification than  waterborne paint, as also indicated by the difference in P\'eclet number.  As a function of $\tau$, waterborne paint changes from $Pe<1$ to $Pe>1$ which causes its drying process to be at the crossover from diffusion to evaporation controlled.

\section{Conclusions}
We have studied the bulk drying process of solventborne and waterborne paints by terahertz time-domain spectroscopy. The measured reflectivity of drying paint layers can be well described by a stratified dispersive model, which provides the thickness and optical properties with high accuracy. The dynamic behavior of the geometrical properties of the coating reveals the interior drying process, which is shown to correlate to the established drying mechanism based on previous work. Moreover, for all paint classes, we observe the growth of a dry top layer on the shrinking wet bulk when the drying is evaporation controlled, consistent with theoretical predictions on evaporating colloidal dispersions.

\end{document}